# The Scaled-Charge Additive Force Field for Amino Acid Based Ionic Liquids

Eudes Eterno Fileti[1] and Vitaly V. Chaban[2]

[1] Instituto de Ciência e Tecnologia, Universidade Federal de São Paulo, 12231-280, São José dos Campos, SP, Brazil

[2] MEMPHYS - Center for Biomembrane Physics, Syddansk Universitet, Odense M, 5230, Kingdom of Denmark

**Abstract**. Ionic liquids (ILs) constitute an emerging field of research. New ILs are continuously introduced involving more and more organic and inorganic ions. Amino acid based ILs (AAILs) represent a specific interest due to their evolutional connection to proteins. We report a new non-polarizable force field (FF) for the eight AAILs comprising 1-ethyl-3-methylimidazolium cation and amino acid anions. The anions were obtained via deprotonation of carboxyl group. Specific cation-anion non-covalent interactions have been taken into account by computing electrostatic potential for ion pairs, in contrast to isolated ions. The van der Waals interactions have been transferred from the CHARMM36 FF with minor modifications. Therefore, compatibility between our parameters and CHARMM36 parameters is preserved. Our FF can be easily implemented using a variety of popular molecular dynamics programs. It will find broad applications in computational investigation of ILs.

**Introduction**

Recent applications of ionic liquids (ILs) have indicated that constitute an important class of materials for use in pure chemistry, chemical engineering and pharmaceutical sciences.[1] Advances in the field during the last decade allow to no longer speak about promising materials, but about a working horse of the modern technology. Due to their unique properties compared to common solvents, ILs have been extensively investigated also in the context of environmental science, since many of them are considered more environmentally friendly than conventional liquids.[1f, 2] Not all ILs are green though. It was shown that certain ILs are toxic and certain ILs emit toxic gases, such as HF and $POF_3$. These gases are produced by the hydrolysis of the fluorine containing anions, such as $BF_4^-$ and $PF_6^-$.[1k]

Since virtually all properties of ionic liquids are adjustable, one can tune them to minimize toxicological effects and environmental impact. The third generation of ILs has emerged in this context. These ILs contain only biocompatible and non-toxic, sometimes bioactive species.. Ionic liquids based on amino acids (AAILs) fall within this category. These liquids were synthesized for the first time about ten years ago and are receiving an ever increased popularity in the research community[2-3] Biocompatibility, in particular, is a highly desirable feature for applications in pharmaceutical sciences, since ionic liquids can be used to synthesize active pharmaceutical ingredients with modified solubility. They can also boost solubility of poorly soluble drugs and drug delivery vehicles.[4]

Recent applications of AAILs feature their value for technology and biomedicine.[2, 3b, 3e, 3g-j, 5] For example, tetrabutylammonium aminoacid-based ionic liquids can be used as green lubricants, indicating that they should be ideal components for future industrial lubricating fluids.[2b] Gas capture, especially $CO_2$, by conventional ionic liquids has been widely studied.[1b, 6] It has recently been shown that imidazolium AAILs may be also used in gas adsorption and separation processes.[2a, 3c, 3k, 5] The last example refers to the use of the AAIL as self-organizing media of amphiphilic molecules. In particular, novel lyotropic liquid crystal systems with the desired behavior can be elaborated.[3h]

*In silico* experiments are of crucial importance for the development of new materials, due to predictive power and relatively low cost. In fact, many of the published work on ionic liquids report computer simulations. Particularly popular are atomistic simulations which use additive mechanistic models of interaction. The success of this technique naturally lies in the definition of the set of parameters and mathematical functions (force field, FF) to be used to model the studied system. Considering the growing interest to AAILs, development of carefully parameterized force fields is highly desirable. The successfully FF will take into account specific cation-anion non-bonded interactions and reproduce electrostatic potential at the surface of ion aggregates. We elaborate a computationally affordable procedure to obtain the FF for 1-ethyl-3-methylimidazolium-based amino acids (emim-AA). Amino acids can be categorized into three different classes according to the electrostatic nature of its side chain (nonpolar, polar non-charged and charged). In this paper, we investigate those AAILs based on hydrophobic side chain amino acid. Therefore, the eight anions considered here are alanine (ala), valine (val), isoleucine (ile), leucine (leu), methionine (met), phenylalanine (phe), tyrosine (tyr) and tryptophan (trp).

**Methodology**

Electronic structure description of AAIL ion pairs was carried out using omega B97XD hybrid density functional theory (DFT) functional.[7] This functional, allegedly, exhibits improved performance for electronic energy levels, non-covalent interactions (importantly) and thermochemistry. Therefore, B97XD was preferred over more traditional functionals. The wave function was expanded using the gaussian basis functions provided in the 6-311G basis set. Additionally, the polarization and diffuse functions have been centered on each heavy atom to provide an improved description of the ions containing high-energy valence electrons (as opposed to neutral molecules). The number of basis functions per system varied with respect to the total number of electrons. No pseudopotentials were used for core electrons. The selected basis set is considered to provide a trustworthy approximation of the real wave function.

The electrostatic potential (ESP) was computed using the Moller-Plesset second-order perturbation theory, MP2, with precisely the same set of basis functions as in omega B97XD. The reason of choosing MP2 over DFT is a stronger compatibility with the CHARMM36 force field.[8] The ESP was approximated using a set of point charges. For simplicity, each charge was centered at an atom, including all hydrogen atoms. The ChelpG scheme with a default grid size in Gaussian 09 was employed to perform a charge assignment.[9] All nuclear geometries (isolated ions and ion pairs) were optimized to correspond to the local minimum configuration of the electron-nuclear system. The electronic structure computations were performed in Gaussian 09, revision D.[10]

We performed three sets of simulations per system per force field. First, enthalpy of vaporization and mass density were determined at room conditions. Second, diffusion constants and ionic conductivities were computed at 450 (500) K using mean-square displacements. Third, non-equilibrium molecular dynamics simulations were performed at 500 K based on continuous energy dissipation in the liquid. The list of the simulated systems is provided in Table 1.

**Table 1**: The list of systems simulated in the present work. The quantity of ion pairs per IL was selected with respect to the cation and anion sizes

| IL | Number of ion pairs | Number of interaction centers | Run 1 (ns) | Run 2 (ns) | Run 3 (ns) |
|---|---|---|---|---|---|
| [emim][ala] | 150 | 4650 | 15 | 50 | 50 |
| [emim][val] | 125 | 4625 | 15 | 50 | 50 |
| [emim][ile] | 100 | 4000 | 15 | 50 | 50 |
| [emim][leu] | 100 | 4000 | 15 | 50 | 50 |
| [emim][met] | 125 | 4750 | 15 | 50 | 50 |
| [emim][phe] | 100 | 4100 | 15 | 50 | 50 |
| [emim][tyr] | 100 | 4200 | 15 | 50 | 50 |
| [emim][trp] | 100 | 4500 | 15 | 50 | 50 |

Cartesian coordinates were saved every 5 ps and thermodynamic quantities were saved every 0.02 ps. More frequent saving of trajectory components was preliminarily tested, but no systematic accuracy improvement was found. Self-diffusion coefficients were computed from mean-square displacements of atomic positions and conductivity was computed from mean-square displacements of translational dipole moment (the Einstein-Helfand fit). Prior to these calculations, the trajectories were pre-processed to remove information about periodic boundary conditions. Shear viscosity was calculated using cosine-shape acceleration of all ions. The mathematical foundation of this method and its applications to molecular dynamics simulations were in detail discussed by Hess.[11] The first nanosecond of the simulation was used for an accelerated ionic flow to be established. The subsequent 19 ns were used for the viscosity calculation.

All systems were simulated in the constant-pressure constant-temperature ensemble. The equations of motion were propagated with a time-step of 2 fs. Such a relatively large time-step was possible due to constraints imposed on the carbon-hydrogen covalent bonds (instead of a harmonic potential). The electrostatic interactions were simulated using direct Coulomb law up to 1.3 nm of separation between the interaction sites. The electrostatic interactions beyond 1.3 nm were accounted for by computationally efficient Particle-Mesh-Ewald (PME) method. It is important to use PME method in case of ionic systems, since electrostatic energy beyond the cut-off usually contributes 40-60% of total electrostatic energy. The Lennard-Jones-12-6 interactions were smoothly brought down to zero from 1.1 to 1.2 nm using the classical shifted force technique. The constant temperature (298 and 500 K) was maintained by the Bussi-Donadio-Parrinello velocity rescaling thermostat[12] (with a time constant of 0.5 ps), which provides a correct velocity distribution for a statistical mechanical ensemble. The constant pressure was maintained by Parrinello-Rahman barostat[13] with a time constant of 4.0 ps and a compressibility constant of 4.5×10$^{-5}$ bar$^{-1}$. All molecular dynamics trajectories were propagated using the GROMACS simulation suite.[14] Analysis of thermodynamics, structure, and transport properties was done using the supplementary utilities distributed with the GROMACS where possible, and the in-home tools.

The thermodynamic (heat of vaporization, $\Delta H_{vap}$, mass density, $d$), structure (radial distribution functions), and transport properties (self-diffusion coefficients, $D_{\pm}$, shear viscosity, $\eta$, ionic conductivity, $\sigma$) have been obtained using conventional molecular dynamics (MD) simulations. The IL ions were placed in cubic periodic MD boxes (Figure 1), whose densities were calculated to correspond to ambient pressure at the requested temperature (298K or 500K).

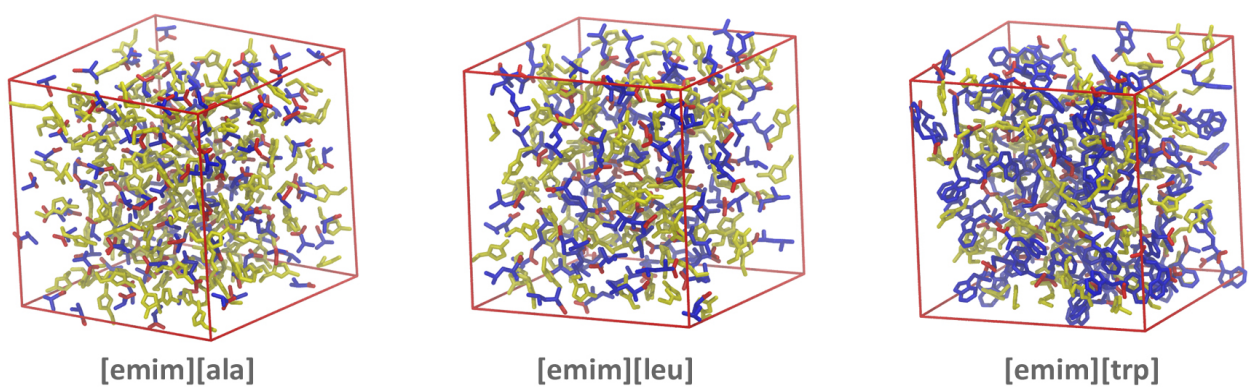

[emim][ala]          [emim][leu]          [emim][trp]

**Figure 1:** Computational cells for three selected amino acid ionic liquids. Imidazolium cation in yellow and amino acid in red (COO¯ moiety) and blue. The abbreviations of anions coincide with the corresponding amino acid.

Many AAILs exhibit extremely high shear viscosities at room temperature. Therefore, simulations have been performed at significantly elevated temperatures to guarantee an appropriate sampling.

**Force Field Derivation**

Figure 2 depicts ion pair configurations corresponding to minimized internal energy. The most positively charged hydrogen atom is coordinated by carboxyl group of all cations. Carboxyl oxygen carries a strong excess electron charge (ESP charge of -0.80e). Its pronounced affinity to an imidazole hydrogen atom can be described as an energetically driven desire to get protonated. Hydrogen bonding is mainly of electrostatic nature. Additionally, the optimized structure of ions is confirmed by classical MD simulations employing the CHARMM36 force field.

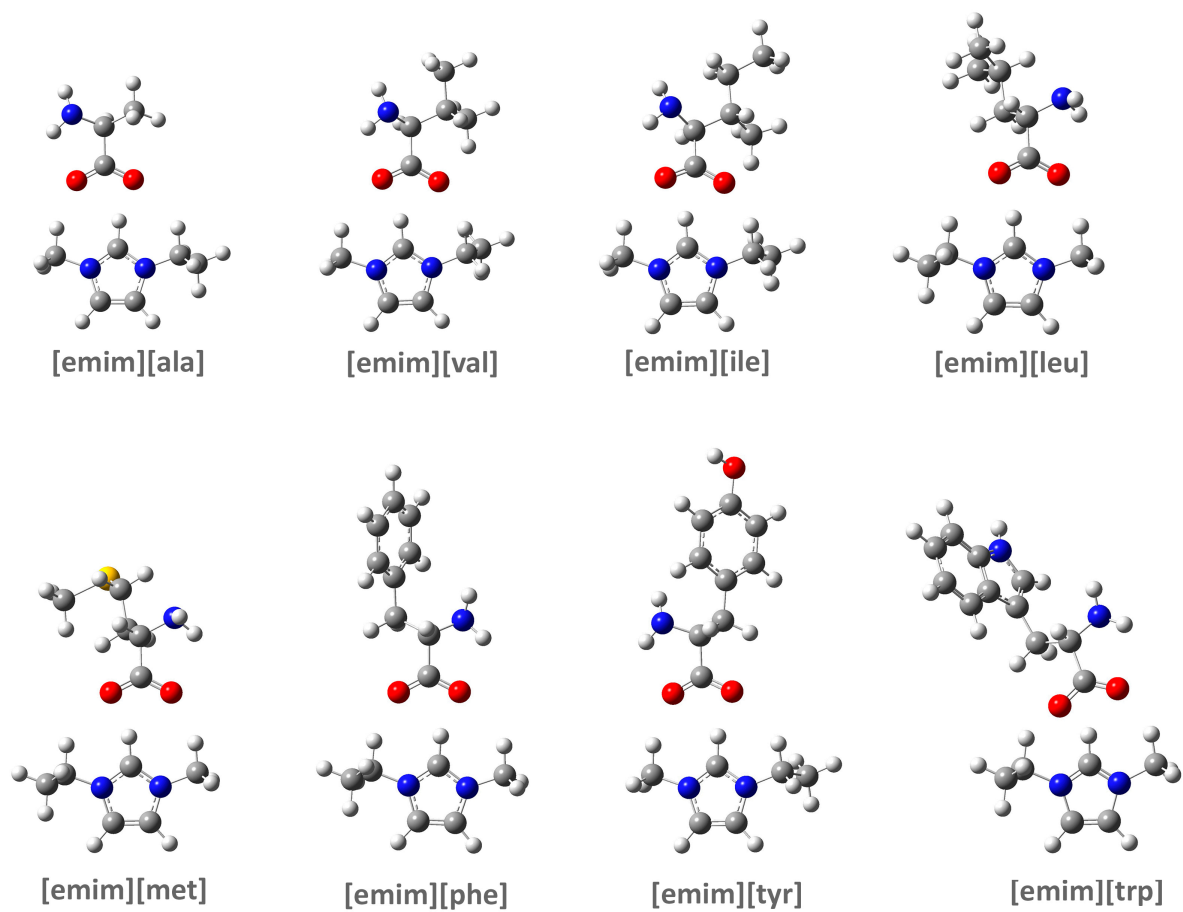

**Figure 2:** Hybrid DFT optimized structures for each of the ion pair investigated. Note, anions are abbreviated using standard notation for the corresponding amino acids.

Well-defined cation-anion coordination implies prevalence of contact ion pairs in all eight AAILs. If these AAILs are dissolved in water or other polar solvent, cation-anion pairs must dominate the solution structure. In turn, solubility in nonpolar solvents and solvents of weak polarity is prohibited by the strong hydrogen bonding. Generally, strong binding of cation or anion is not favorable for ionic liquids, since it increases melting temperature. Industrial applications of this sort of AAILs will be most successful in combination with low-viscous polar solvents.

According to hybrid DFT, electronic polarization takes place between the COO⁻ moiety of an anion and imidazole ring of the cation. This effect brings non-additivity into non-bonded interaction potential. Therefore, CHARMM36 parameters cannot describe these systems perfectly. Importantly, the difference between various amino acid anions, in terms of assigned ESP point charges (Table 2), does not exceed the expected difference for various anion conformations during molecular dynamics at finite temperature. Due to this feature, the refinement procedure can be unified to account only for carboxyl group of the anion. Amino group is not sensitive to the presence of imidazole ring.

We selected [emim][ala] ion pair to derive point electrostatic charges for the corresponding moieties of all eight AAILs. The ion pair geometry was reoptimized using MP2 method and ESP potential was fitted on the basis of the highly accurate MP2 electron density. The obtained charges amount to -1.17e (nitrogen of amino group), +0.38e (hydrogen of amino group), -0.80e (oxygen of carboxyl group), +0.75e (α-carbon). α-carbon is a "buried" atom in the considered configuration. That is, this atom is not located at the surface. Point charge assignment for such atoms is usually ambiguous. We will not use the charge suggested by ChelpG procedure, but assign the charge to neutralize an ion pair.

**Table 2**: Analysis of point charges distribution obtained by fitting the electrostatic potential for the eight amino acid based ILs. Electrostatic point charges are given in electron units. The geometries of all ion pairs were optimized using omega B97XD DFT method.

|       |          | q[O(COO)] | q [C(COO)] | q [$C_\alpha$] | q [N($NH_2$)] | q[H($NH_2$)] |
|-------|----------|-----------|------------|----------------|---------------|--------------|
| [ala] | in pair  | -0.78     | +0.71      | +0.65          | -1.12         | +0.37        |
|       | isolated | -0.84     | +0.77      | +0.54          | -1.11         | +0.35        |
| [val] | in pair  | -0.74     | +0.76      | +0.24          | -1.10         | +0.39        |
|       | isolated | -0.84     | +0.87      | +0.24          | -1.14         | +0.40        |
| [ile] | in pair  | -0.74     | +0.81      | +0.18          | -1.08         | +0.38        |
|       | isolated | -0.83     | +0.86      | +0.28          | -1.15         | +0.38        |
| [leu] | in pair  | -0.82     | +0.76      | +0.41          | -1.05         | +0.36        |
|       | isolated | -0.84     | +0.84      | +0.35          | -1.06         | +0.35        |
| [met] | in pair  | -0.83     | +0.75      | +0.38          | -1.07         | +0.37        |

|       | isolated | -0.83 | +0.83 | +0.30 | -1.09 | +0.36 |
|-------|----------|-------|-------|-------|-------|-------|
| [phe] | in pair  | -0.76 | +0.69 | +0.67 | -1.06 | +0.35 |
|       | isolated | -0.79 | +0.69 | +0.75 | -1.01 | +0.30 |
| [trp] | in pair  | -0.77 | +0.68 | +0.62 | -1.14 | +0.37 |
|       | isolated | -0.78 | +0.64 | +0.73 | -1.12 | +0.33 |
| [tyr] | in pair  | -0.77 | +0.68 | +0.67 | -1.06 | +0.34 |
|       | isolated | -0.79 | +0.69 | +0.69 | -1.00 | +0.30 |

The ESP charge located on the imidazole moiety in the ion pair configuration equals to +0.85e (in contrast to +1e in the isolated ion configuration). The charge deficiency is delocalized over imidazole ring and cannot be ascribed to any particular interaction center. For this reason, we uniformly scale down imidazole point charges. The original charges on emim$^+$ were obtained for the isolated cation to preserve isolated cation symmetry. These charges were multiplied by 0.85 and rounded to two decimal digits. The ions pair was neutralized by manually setting up the charge on the α-carbon (-0.32e). The correctness of our methodological choice is confirmed by Table 2, which lists very different charges (ChelpG scheme) on this atom in various AAIL ion pairs. No other polar atom exhibits such physically irrelevant fluctuations. To recapitulate, the assignment was done for the polar moieties of AA anions. The Coulombic charge set is, therefore, transferrable.

The CHARMM36 FF slightly overestimates hydrogen bond distance between carboxyl group oxygen atom and hydrogen atom of imidazole ring. The cross-sigma parameter for this interaction was set to 0.16 nm in accordance with the ion pair structure optimized using hybrid DFT functional, omega B97XD.

**Results and Discussion**

Figure 3 shows the radial distribution functions between the most charged sites of each ion, namely the H atom located between nitrogen atoms of the imidazolium ring and oxygen atoms of amino acid COO$^-$ group. A high sharp peak is observed at ~0.17 nm. Interestingly, this feature is common for all eight investigated AAILs. Such strong cation-anion coordination is unfavorable for room-temperature ionic liquids, since it fosters crystallization at higher temperature. Therefore, large viscosity must be expected. Furthermore, the first peak suggests a stability of ion pairs in the condensed phase of pure ILs and in mixtures with molecular cosolvents. This is in contrast to fully solvated ions, solvent separated cation-anion pairs, and larger aggregates. The second peak is located at ~0.6 nm. It is small and poorly localized. Such a peak confirms inability of AAILs to create large electrostatically driven aggregates, as opposed to stable ion pairs.

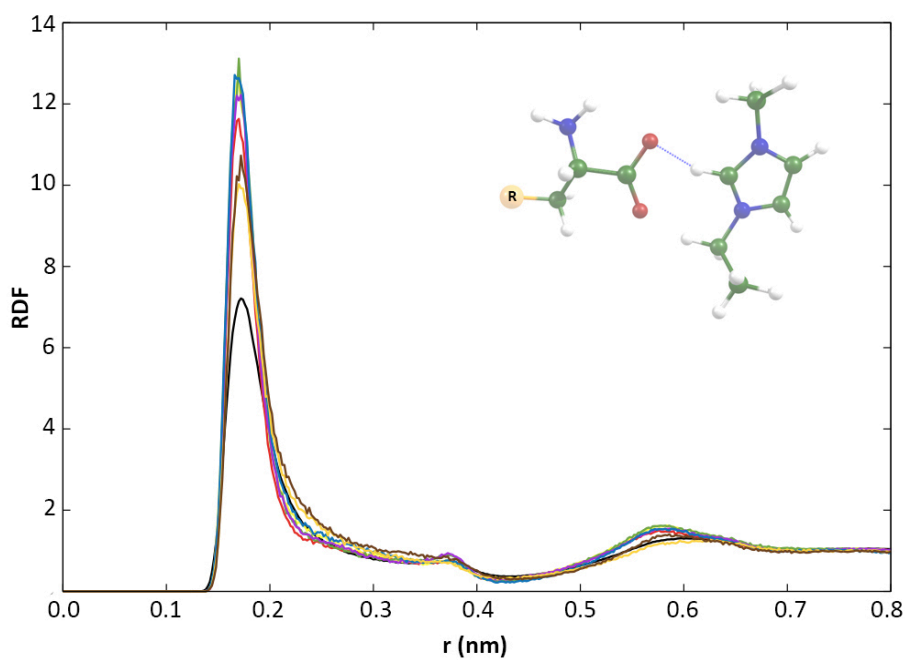

**Figure 3:** Radial distribution functions computed for the most positively charged atom of the cation (ring hydrogen) and the most negatively charged atom of the anion (O from COO¯ amino acid moiety). The aail-color codes are: ala-black, val-red, ile-yellow, leu-green, met-purple, phe-blue, tyr-orange and trp-brown. Inset: the most energetically favorable cation-anion coordination.

Tables 3 and 4 summarize thermodynamic (density, heat of vaporization) and transport (diffusion, viscosity, conductivity) properties obtained using the original CHARMM36 models and our force field for AAILs.. Densities and enthalpies were obtained from simulations at room temperature (298 K). The results for the new model differ insignificantly from the original model. For instance, the discrepancies in mass density do not exceed 1%. The discrepancy in enthalpy of vaporization is larger, up to 15%. Our DFT-driven refinement decreases interaction strength, and consequently, decreases enthalpy of vaporization.

The density of AAILs ranges from 1058 kg m$^{-3}$ ([emim][leu]) to 1187 kg m$^{-3}$ ([emim][trp]). It is somewhat higher than water density at the same conditions. In the meantime, density is smaller than of more traditional [emim] containing ILs, such as [emim] tetrafluoroborate and [emim] bis(trifluoromethanesulfonyl)imide.[15] Density comparison is important, because most currently envisioned applications of AAILs will probably deal with water containing environments. A good or excellent solubility of AAILs in water is anticipated due to a hydrophilic amino acid fragment. The experimental data on AAILs are scarce yet. To our best knowledge, only density of [emim][ala] is known, being 1121 kg m$^{-3}$. The simulated density is 1140 kg m$^{-3}$, which is in a good agreement with the experimental value. The discrepancy (1.7%) is roughly comparable to thermal fluctuations of volume in the simulated systems.

Table 3: Properties of amino acid based ionic liquids obtained using theCHARMM36 FF at 450 K. .

| Liquid | ρ (kg m$^{-3}$) | ΔH$_{vap}$ (kJ mol$^{-1}$) | D(+) (μm$^2$ s$^{-1}$) | D(−) (μm$^2$ s$^{-1}$) | η (cP) | σ (mS cm$^{-1}$) |
|---|---|---|---|---|---|---|
| [emim][ala] | 1147 | 184 | 13±1 | 8±1 | 4.6±0.1 | 160±133 |
| [emim][val] | 1089 | 182 | 7±0 | 4±0 | 6.6±0.2 | 94±85 |
| [emim][ile] | 1069 | 184 | 6±1 | 3±0 | 6.6±0.3 | 64±38 |
| [emim][leu] | 1063 | 188 | 7±1 | 4±0 | 5.6±0.2 | 54±23 |
| [emim][met] | 1158 | 201 | 6±1 | 4±0 | 6.5±0.1 | 34±27 |
| [emim][phe] | 1145 | 204 | 5±0 | 2±0 | 7.7±0.4 | 71±38 |
| [emim][tyr] | 1183 | 236 | 0.4±0.1 | 0.2±0.1 | 76.0±13.0 | 4±2 |
| [emim][trp] | 1175 | 227 | 0.2±0.1 | 0.1±0.0 | 100±20 | 7±3 |

Table 4: Properties of amino acid based ionic liquids obtained using the newly developed force field at 500 K. .

| Liquid | ρ (kg m$^{-3}$) | ΔH$_{vap}$ (kJ mol$^{-1}$) | D(+) (μm$^2$ s$^{-1}$) | D(−) (μm$^2$ s$^{-1}$) | η (cP) | σ (mS cm$^{-1}$) |
|---|---|---|---|---|---|---|
| [emim][ala] | 1140 | 157 | 70±2 | 50±5 | 2.2±0.0 | 690±250 |
| [emim] | 1082 | 154 | 59±9 | 35±4 | 2.7±0.1 | 455±152 |
| [val] | | | | | | |
| [emim][ile] | 1064 | 157 | 48±7 | 31±3 | 2.8±0.1 | 203±246 |
| [emim][leu] | 1058 | 160 | 57±7 | 35±2 | 2.5±0.1 | 432±272 |
| [emim][met] | 1152 | 174 | 47±4 | 29±2 | 3.1±0.0 | 259±140 |
| [emim][phe] | 1144 | 176 | 40±3 | 24±2 | 3.7±0.1 | 262±152 |
| [emim][tyr] | 1182 | 217 | 10±1 | 2±1 | 36.3±4.2 | 78±53 |
| [emim][trp] | 1187 | 208 | 6±1 | 2±0 | 34.5±2.2 | 58±36 |

The ionic transport (self-diffusion constant, ionic conductivity) significantly increases when electronic polarization effects are attended. Compare Tables 3 and 4. In turn, shear viscosity decreases proportionally. Viscosity varies greatly depending on the anion. [tyr] and [trp] featuring hydrophilic radicals are notably viscous, 36 and 35 cP at 500 K, respectively. In turn, [ala], [val], [leu] possessing small radicals are not viscous. Large hydrophobic radical, such as in [phe], makes a minor impact on the transport properties, as compared to smaller hydrophobic radicals. Unfortunately, no experimental reports on shear viscosity are still available.

[emim] cation is generally more mobile than any AA anion. However, an absolute value depends strongly on the anionic species. This is an additional indication of strong cation-anion binding in this class of ionic liquids. Diffusion of cation in [emim][trp] is more than ten times slower than in [emim][ala]. Diffusion of anion is more than twenty times slower. We conclude that amino acid ILs exhibit very different transport properties, although the structure of all ion pairs is similar. Such diversity opens wide opportunities to tune properties by mixing AAILs with one another and with polar molecular cosolvents. High viscosity at lower temperatures suggests that usage of pure AAILs for applications is unlikely. An exception is [emim][ala], because of its small hydrocarbon radical. Other AAILs may be more successful in combination with water; for instance, to solubilize biologically relevant peptides and proteins.

**Concluding Remarks**

We have developed a new force field for the eight imidazolium-based amino acid ionic liquids. The FF does not contain an explicit interaction term responsible for electronic polarization effects, which results in a higher computational affordability. Specific cation-anion interactions, including polarization, are on the average captured by the refined electrostatic potential. Unlike in the previous FFs, electrostatic potential has been derived using electronic structure description of a neutral ion pair. The developed force field fosters computer simulations of pure AAILs and their mixtures with biologically relevant agents.

Note, there are several procedures to fit ESP using point charges. Different procedures provide somewhat different charge distributions. In turn, different electronic structure methods coupled with different basis sets impact the ESP to be fitted. One should not expect that the scaling factors derived in the present work are absolute quantities. Additionally, point ESP charges and charge transfer effects are conformation dependent. Only a single local-minimum conformation of each IL was considered for the charge refinement in our procedure. The results indicate that our approximation provides reasonable accuracy in the light of expectations to phenomenological models.

Extension of the proposed methodology to other representatives of AAILs is underway.

**Supporting Information**

The force field derived in this work and ready-to-use topology input files for the GROMACS molecular dynamics simulation package are available free of charge upon e-mail to

fileti@gmail.com or vvchaban@gmail.com. Please, include your name and academic affiliation in the message.

## AUTHOR INFORMATION

E-mail addresses for correspondence: fileti@gmail.com (E.E.F.); vvchaban@gmail.com; chaban@sdu.dk (V.V.C.)